\documentclass[12pt,preprint]{aastex}

\usepackage{color}

\usepackage[normalem]{ulem}

\shorttitle{Kelvin-Helmholtz instability in a streamer}

\shortauthors{Feng et al.}

\begin{document}

\title{Kelvin-Helmholtz instability of a coronal streamer}

\author{L. {Feng}\altaffilmark{1,2}, B. {Inhester}\altaffilmark{2}, W.Q. {Gan}\altaffilmark{1}}

\email{lfeng@pmo.ac.cn}

\altaffiltext{1}{Key Laboratory of Dark Matter and Space Astronomy, 
  Purple Mountain Observatory, Chinese Academy of Sciences, 210008 Nanjing, China}

\altaffiltext{2}{Max-Planck-Institut f\"{u}r Sonnensystemforschung, Max-Planck-Str.2,
37191 Katlenburg-Lindau, Germany}

\begin{abstract}

The shear-flow-driven instability can play an important role in energy transfer 
processes in coronal plasma. We present for the first time the observation of
a kink-like oscillation of a streamer probably caused by the streaming
kink-mode Kelvin-Helmholtz instability. 
The wave-like behavior of the streamer was observed by Large Angle and Spectrometric 
Coronagraph Experiment (LASCO) C2 and C3 aboard 
{\it{SOlar and Heliospheric Observatory}} (SOHO).
The observed wave had a period of about 70 to 80 minutes, and its wavelength increased from  
$2\;R_\odot$ to $3\;R_\odot$ in about 1.5 hours. The phase speeds of
its crests and troughs decreased from $406\pm20$ to $356\pm31 \mathrm{km s^{-1}}$
during the event. Within the same heliocentric
range, the wave amplitude also appeared to increase with time. We attribute the phenomena to
the MHD Kelvin-Helmholtz instability which occur at a neutral sheet in a fluid wake.
The free energy driving the instability is supplied by the sheared flow and sheared magnetic
field across the streamer plane.
The plasma properties of the local environment of the streamer were estimated
from the phase speed and instability threshold criteria.

\end{abstract}


\section{Introduction}

The Kelvin-Helmholtz instability (KHI) driven by a shear of the flow velocity,
has been observed in various astrophysical environments. In the solar
atmosphere, \citet{Berger:etal:2010} and \citet{Ryutova:etal:2010} discovered
it in a prominence from {\it Hinode} Solar Optical Telescope observations.
\citet{Foullon:etal:2011} and \citet{Mostl:etal:2013} found that it could
develop at the flank of a fast coronal mass ejection.
\citet{Ofman:Thompson:2011} identified a vortex-shaped feature caused by the
KHI on the boundary of a dimming area during the eruption of an active region.
It has been shown by \citet{Chandrasekhar:1961} that the presence of a
magnetic field tangential to the boundary surface and directed along the flow
stabilizes the hydrodynamic KHI. For a symmetric shear layer, the instability
can only grow if the the velocity jump across the boundary between the two
uniform regions exceeds twice the Alfv\'en velocity on either side.

The instability growth is modified when the field and
density profiles across the boundary are asymmetric or if the shear layer has
a finite width. Instability calculations and simulations have been performed
to model the magnetospheric current sheet, astrophysical jets and
streamer sheets in the solar corona. The model adopted for these cases were
2D and 3D force-free or pressure balanced current sheets with a jet or a wake
layer centered at the current sheet
\citep{Lee:etal:1988,Wang:etal:1988,Dahlburg:etal:1998, Einaudi:1999,
  Dahlburg:etal:2001, Zaliznyak:etal:2003}.
Owing to Galilean invariance, the wake, a layer of reduced speed embedded
in a uniform background flow, is equivalent to the jet case for otherwise
identical plasma parameters and produces the same instability growth rates 
but shifted phase speeds \citep{Einaudi:etal:1999}.
More recently, 2.5~D and 3~D simulations were performed to shed more light on
the velocity and magnetic shear driven instabilities in the nonlinear regimes
\citep{Bettarini:etal:2006, Bettarini:etal:2009}.

From all these studies, three different modes could be distinguished during
the initial phase of the instability. For low Alfv\'enic Mach numbers below
about two, the instability is strongly influenced by the magnetic field and
shows plasmoid formation as a result of the tearing mode instability (or
resistive varicose mode) \citep[e.g.,][]{Einaudi:etal:1999}. For Alfv\'enic
Mach numbers above about two, the hydrodynamic behavior overwhelms the
magnetohydrodynamic evolution. A symmetric and an asymmetric oscillatory mode
evolve in such a situation termed kink (or sinuous) mode and sausage
(or ideal varicose) mode.

Using the linear theory, \citet{Lee:etal:1988} derived analytical expressions
for the growth rates and phase speeds of the three modes of a jet assuming a four layer
model each with constant plasma parameters. They found that the kink mode
dominates over the sausage mode when the value of $\beta$ (ratio of thermal to
magnetic pressure) in the ambient plasma exceeds unity. This value of $\beta$
is, however, linked to other equilibrium parameters of the current sheet by
the requirement to maintain the net pressure balance, e.g., the
ratio of the density between the current sheet and its ambient plasma. 
In an accompanying paper, \citet{Wang:etal:1988} applied the wake model to the heliospheric
current sheet and considered the influence of difference 
between the width of the velocity shear layer and the width of the current 
sheet on the growth rate. Numerical simulations of the
current carrying wake with various plasma parameters suggest that the kink
mode often dominates for the large $\beta$ parameters considered, except for
long wavelengths of the perturbations \citep{Dahlburg:etal:1998,
Dahlburg:etal:2001,Zaliznyak:etal:2003}. Even though the instability growth
rate decreases steadily with increasing sonic Mach number, i.e., with
decreasing $\beta$, the simulations do not show a drastic threshold for
$\beta$ or for the sonic Mach number like is exists for the Alfv\'enic Mach
number.

The quasi 2D current plane at the tip of a helmet streamer seems to become
particularly close to the model geometry adopted in the above simulations.
However, contrary to the ease with which the kink mode is excited in the
simulations, it is not observed as frequently. Indeed, SOHO/LASCO observations
show the almost continuous ejection of plasmoids from the tip of helmet
streamers, \citep[e.g.,][]{Sheeley:etal:1997}. They were interpreted as the
development of the tearing-type reconnection to a nonlinear varicose
(sausage-like) fluid instability discussed above
\citep{Einaudi:etal:1999,Einaudi:etal:2001}. On the other hand, the equivalent
kink-like instability along a streamer has to the knowledge of the authors not
been observed.

In this paper we report the rare observation of kink-like oscillations in the
current sheet of a coronal streamer. On June 3, 2011 a narrow streamer was
emerging from behind the occulter of SOHO/LASCO C2 and C3 which after about
7:30 exhibited a clear oscillating behavior.

A wave-like motion of a streamer was also observed by \citet{Chen:etal:2010}
which the authors claim to have been excited by the eruption of a fast coronal
mass ejection (CME) nearby. In their observations, the wave
was excited impulsively and steadily decayed afterwards.
In the case of the streamer on June 3 reported
here, no CME occurred close to the streamer, neither in time nor in space.
Also, the wave amplitude in this event stayed constant for
hours or even slightly increased with time. We therefore attribute this oscillation to
an instability rather than to an impulsive excitation.
To our knowledge it is the first observation of a streamer wave which was
probably caused by the streaming kink instability. In Section 2, the
observations of the streamer wave by the LASCO coronagraph and the retrieving
of the wave shape are described. In Section 3, we compare the potential of the
streaming kink instability and other competing MHD modes of the KHI as a source
for this wave. In the last section, the obtained results will be summarized
and discussed.

\section{Observations and data reduction}

\begin{figure} 
\centering
\includegraphics[width=15.cm, height=15.cm]{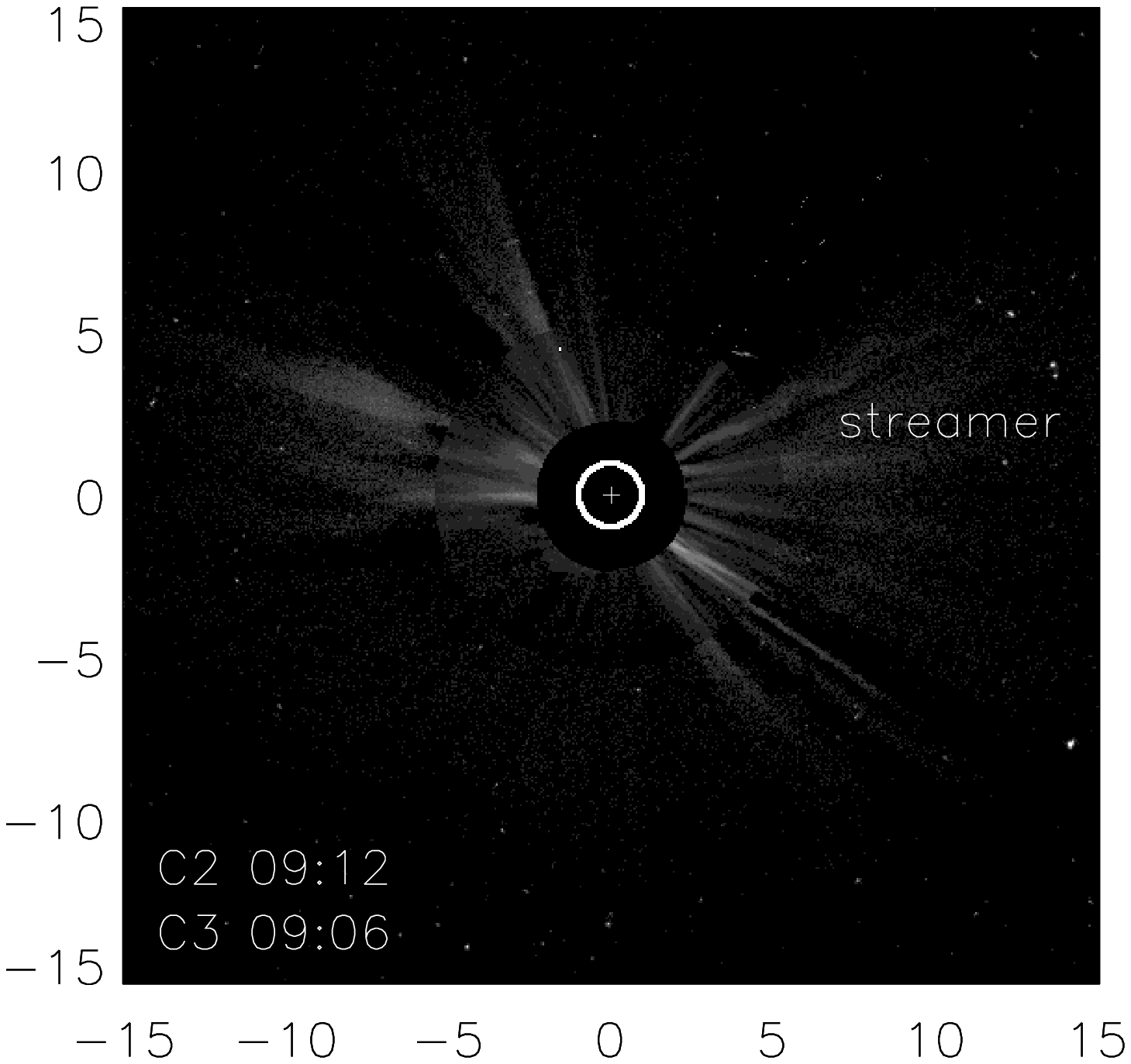}
\caption{The oscillating streamer observed by C2 at 09:12~UT and observed by 
C3 at 09:06~UT. The white circle indicates the solar limb, and the plus sign 
represents the Sun center. A movie showing the evolution of the oscillating 
steamer is available online. The units are in solar radius.}
\label{fig:c2c3combine}
\end{figure}

The oscillating feature was observed by the LASCO \citep{Brueckner:etal:1995}
C2 and C3 coronagraphs aboard SOHO. LASCO C2 has a FOV from 2 to
6~R$_{\odot}$, C3 from 3.7 to $30\;R_{\odot}$. The wave-like motion of the
streamer first appeared in the C2 field of view at a distance 3 to $4\;R_\odot$
at 05:48~UT on June 3, 2011. The wave which developed then
propagated almost radially to a distance of about $11 \;R_{\odot}$ at
10:30~UT. At larger distances, the phenomenon becomes 
impossible to observe due to the low signal to noise.
A movie showing the streamer wave can be found in the online
material. The end of the wave-like motion was accompanied
by a disruptive eruption beginning around 13:25~UT.
In Figure~\ref{fig:c2c3combine}, a snapshot of the oscillation is
presented. The streamer with the wave-like feature is located in
the first quadrant of the image plane and is rooted at a latitude
around 27.5$^{\circ}$. 

From the LASCO images, the streaming feature appears as a white-light jet
\citep{Wang:etal:1998, Feng:etal:2012}. Usually, a jet can be traced down to
its footpoints at the solar surface. However, AIA/SDO did not observe any
possible source for the jet at lower altitude. Moreover, an
elongated structure existed at the same position for many hours before the 
appearance of the wave. An alternative possibility is
the leakage of plasma from the cusped magnetic field lines above a helmet
streamer. The cusp is a hot region at the streamer top with low magnetic field
strength and the magnetic boundary concavely bent outwards. In this region the
plasma $\beta$ may easily have values above unity. Thus the magnetic field
confinement can be considered marginal and the streamer plasma may
occasionally leak out of the cusp region to feed the slow solar wind. It is
generally observed that the cusp altitude is below 2 to $2.5\;R_{\odot}$, very
close to the inner boundary of the LASCO C2 FOV \citep{Chen:etal:2010}. In EUV
observations, due to the relatively low density of the helmet streamer plasma
and the restricted field of view, it is usually difficult to see the dome part
of a streamer. For that reason, AIA was probably not able to see the lower
helmet counterpart associated with the observed streamer.

To clarify the spatial context of this event, we show an EUV image observed
at $\lambda=28.4$~nm by the ahead spacecraft of \textit{Solar TErrestrial RElations Observatory} 
\citep[STEREO,][]{Kaiser:etal:2008}
in the upper panel of Figure~\ref{fig:spatial_context}. At the time of 
observation, STEREO A was located about 90 degrees west of the Sun-Earth line
and looks almost vertically down onto the longitude of the streamer. 
Two coronal holes, one above and one below the streamer latitude are
indicated.

The lower panel of Figure~\ref{fig:spatial_context} shows the field lines
from a PFSS model emerging from a $20^{\circ}$ wide longitude range around the
limb and from latitudes of $27.5\pm30^{\circ}$. The field extrapolation was
produced from a synodic magnetogram observed by the Helioseismic and Magnetic
Imager \citep[HMI;][]{Schou:etal:2012} aboard {\it Solar Dynamic
Observatory} (SDO) for Carrington rotation (CR) 2110. In this magnetogram, the
limb longitude is intersected by two neutral lines which for a source surface
height $r_{SS}<1.6 R_\odot$ results in two current sheets above the limb. The
southern sheet matches very closely the latitude of observed
oscillating streamer. For larger $r_{SS}$, the two current sheets merge and
the PFSS field beyond the source surface height becomes unipolar. In this
case, the observed streamer would have to be interpreted as a pseudostreamer
\citet{Wang:etal:2007}.
From the observations in Fig.~\ref{fig:c2c3combine}, we can state that the
source surface height $r_{SS}$ for the oscillating streamer must be well
less than the LASCO C2 occulter radius of about 2.2 $R_\odot$. The fact that
a value of $r_{SS}<1.6 R_\odot$ produces a current sheet within a few degrees
of the observed streamer latitude gives us confidence that the observed
streamer is associated with a current sheet and the streamer bulge is
confined to low altitudes.

In Figure~\ref{fig:c2c3trace}, some selected image frames which zoom into the
streamer wave are presented. To make the streamer structure more prominent, a
pre-event image immediately before the streamer's appearance was subtracted
from the following sequence of C2 and C3 images, respectively. For a better
visualization of the wave-like structure and its propagation, the streamer was
rotated clockwise by 27.5$^{\circ}$ to horizontal. The shape of the streamer
axis was traced by hand and marked by red plus signs. In each frame, if
applicable, the crest and trough are indicated in yellow. The first crest is
enumerated by P1, the first trough by P2, and the second crest by P3. From the
C2 observations, the streamer started to oscillate only above the distance of
about $3\;R_\odot$ from the Sun center. After 08:12~UT, the streamer
brightness intensified, which can be seen, e.g., from the large brightness
gradient along the streamer near $4.6\;R_{\odot}$ at 09:24~UT. This also
modified the shape of the streamer especially below the heliocentric distance
of 5 to $6\;R_{\odot}$.

A complete dataset of the traced streamer axis from C2 and C3
images is displayed in Figure~\ref{fig:c2c3all}. The time cadences of the C2
and C3 images were both 12 minutes in general but occasionally had an
observational gap of more than 12 minutes. The streamer traced in C2 is
delineated with dashed lines, in C3 with solid lines. Streamer
traces observed with a time difference of less than 6 minutes are drawn in the
same color. Although after 08:12~UT, the full length of the streamer was
traced out, we will neglect in the further analysis the segment behind the
brightness intensification noted above.

The traced streamer positions in Figure~\ref{fig:c2c3all} may be subject to
some uncertainty. To investigate the amplitude variation with distance and time, the
streamer segments around the first and second crests P1 and P3 are compared in
Figure~\ref{fig:amplitude} after they propagated for somewhat more than an
hour from the C2 field of view into C3. The upper dashed horizontal line in
Figure~\ref{fig:amplitude} marks the peak of P3, the lower dashed line the
peak of P1. It shows that the upper boundary of the crests P1 
and P3 increased from 0.3 to $0.5\;R_{\odot}$ when they propagated to larger 
distances from about 5.2 to $7.5\;R_{\odot}$, which indicates the increase of their
amplitudes with distance.
If we use the two short solid lines as the lower boundary of the
streamer segment around its peak at 07:36~UT and 08:48~UT, the amplitude, as
calculated from half the average distance of the lower and upper boundaries,
increases slightly from 0.14 to $0.17\;R_{\odot}$. It also increases
slightly with time during our observations.
Although these estimates of the amplitude increase are not very precise due to data noise, it is very
unlikely that the streamer in our study experienced an amplitude decrease
as it was observed by \citet{Chen:etal:2010}. Therefore, we rule out
that the wave-like behavior of the streamer in our observations was caused by
an interaction of the streamer with a CME as reported in their paper.

Since the brightness in LASCO C2 and C3 is caused by Thomson scattering of
electrons in the corona, the observed coronagraph brightness is proportional
to the column density along the line-of-sight. Moreover, we conclude form the
fact that the streamer can distinctly be seen on the dark background sky and
appears like a narrow ribbon that we observe the streamer more or less
edge-on. We therefore assume that the observed column densities can serve as a
proxy of the volume density along a profile normal to the streamer axis. In
the upper panel of Figure~\ref{fig:ROI_den}, the eight red lines mark the
profiles along which the brightness measurements were performed. Note that the
pre-event image was not subtracted from this image. Instead, a monthly-minimal
image was subtracted to remove the stray light and/or the F corona. 

In Figure~\ref{fig:den_ratio}, we display the brightness distribution across
the streamer on these eight profiles, extracted from the observation at
07:24~UT when the oscillation of the streamer started to become pronounced.
Two dotted lines represent the maximum $\rho_s$ and the assumed background
$\rho_\infty$ of the brightness. Obviously, the ratio $\rho_s/\rho_\infty$
declines approximately with heliocentric distance from $\rho_s/\rho_\infty\simeq 2$ to
$\rho_s/\rho_\infty\simeq 1.5$ except at $r=4 \;R_\odot$. The brightness distribution at
$r=2.5\;R_\odot$ was contaminated by a bright structure in the south of the
streamer. We therefore used for this heliocentric distance the background estimate from
$r=2.8 \;R_\odot$ instead. Interestingly, the streamer wave is discernible
only at a distance beyond $r=3.5\;R_\odot$ when
the density ratio has already settled to its lower bound of
$\rho_s/\rho_\infty\simeq 1.5$. As another parameter, we can estimate from
Figure~\ref{fig:den_ratio} the width of the current sheet. For altitudes below
$r=4 \;R_\odot$ the streamer is well concentrated with a half-width of
$\delta\simeq 0.2 \;R_\odot$. Beyond $4 \;R_\odot$ the streamer 
appears more diffusive probably due to decreasing signal-to-noise.

There are two major sources of uncertainties involved in 
the estimate of the density ratio. One is the unknown line-of-sight depth of
the streamer. We assumed the density ratio to be equal the ratio of the
observed column densities which implies an infinite streamer depth. 
If the real depth of the streamer is well less than $R_\odot$, we underestimate the
density ratio.
The other unknown is related to the background subtraction of the coronagraph
image. To obtain an upper bound of the ratio, instead of the monthly-minimum
background, a yearly-minimal image was utilized. The corresponding density
ratios are indicated in the parentheses in each panel. Although the ratios
increase slightly, the trend of decreasing ratio with distance is clear.

Based on the retrieved wave-like structure in Figure~\ref{fig:c2c3all}, we
have estimated parameters which characterize the streamer wave. The
heliocentric distance of the crests and troughs, P1, P2 and P3, is plotted in
Figure~\ref{fig:speed} as a function of time. Applying linear fits to their
space-time positions yields a phase speed for P1, P2 and P3 of
$406\;\mathrm{km s^{-1}}$, $374\;\mathrm{km s^{-1}}$ and $356\;\mathrm{km
s^{-1}}$, respectively. The decrease of the phase speed from P1 to P3 can be
directly concluded from the slight divergence of the fit lines in
Figure~\ref{fig:speed}. As seen in Figure~\ref{fig:c2c3trace}, at 08:06~UT the
half wavelength $\overline{P1P2}$ was about $1\;R_\odot$; by 09:30~UT the half
wavelength $\overline{P1P2}$ had enlarged to $1.5\;R_\odot$.
The period of the wave was measured from the time interval between the first
to the second crest, P1 and P3. We found about a period of 72 minutes at $r=5
\;R_\odot$ and 84 minutes at $r=9 \;R_\odot$.

The major uncertainty involved in the above analyses comes from the
delineation of the streamer profile in the LASCO observations
(Figure~\ref{fig:c2c3trace}). Once the profiles are determined, tracking the
crests and troughs of the profiles and fitting a linear function to their
distance-time plots only contributes to a minor uncertainty. We assume that
the uncertainty of the streamer identification corresponds to the half width
$\delta$ of the streamer of about $\pm 0.2 \;R_\odot$. This is also the
uncertainty of the positions of the crests and troughs in
Figure~\ref{fig:speed}. The linear fits bearing such uncertainties produce a
1-sigma error of the three velocity estimates of 16, 9 and $24\;\mathrm{km
s^{-1}}$.

\section{The streaming kink instability}

Analytical and MHD simulations have shown that streaming sausage, kink and
tearing instabilities can occur in presence of sheared velocity and magnetic
field systems. At the tip of a helmet streamer, the magnetic field has
opposite directions on the two sides with a neutral sheet inside. The field
strength inside the streamer is probably well below the value outside
and the pressure deficit is compensated by an enhanced plasma density.
The plasma velocity assumes a minimum in the neutral current sheet which
can be considered the source of the slow solar wind
and increases towards both sides where the dilute plasma is accelerated
into the fast solar wind.
Plasma blobs probably formed by the streaming sausage and tearing
instabilities and convected away by the surrounding solar wind flow have been
observed at the tip of the helmet streamer \citep[e.g.,][]{Sheeley:etal:1997}.
So far however, kink mode waves have not been detected yet. In view of the
amplitude increase of the wave-like structure we observe, we tend to believe
that it is instability-driven. The increase of the perturbation is probably
fed from the kinetic energy of the surrounding solar wind plasma.

A very robust threshold for the streaming kink and sausage instability found
in many simulations concerns the velocity difference of the shear flow. In 
the simulation of
\citet{Chen:etal:2009}, an Alfv\'en speed of $20\;\mathrm{km s^{-1}}$ inside
and of $75\;\mathrm{km s^{-1}}$ outside of the current sheet was assumed. With
these parameters \citet{Chen:etal:2009} could excite both modes with a
velocity difference for the shear flow of $100\;\mathrm{km s^{-1}}$ for the
kink and about $150\;\mathrm{km s^{-1}}$ for the sausage mode.

If we take the measurements by the Ulysses space craft
\citep{McComas:etal:2008} as a typical estimate of the solar wind speed and
its gradients, these threshold values for the velocity shear can be easily
achieved. Note, however, that the Ulysses measurement were made beyond 1 AU
where the solar wind has been accelerated to its asymptotic speed which
probably has not yet been reached at 10 $\;R_\odot$. Also the Alfv\'en speed
assumed by \citet{Chen:etal:2009} is relatively low. A typical value beyond
$2\;R_\odot$ is $200\;\mathrm{km s^{-1}}$ \citep[e.g.,][]{Warmuth:Mann:2005}.
However, in cusp and streamer regions the field strength is probably
reduced with respect to this average.

\citet{Lee:etal:1988} found from an analytic solution of a four-layer model
that the kink mode dominates over the sausage mode when $\beta$ in the ambient
plasma exceeds unity, or equivalently, $\rho_s/\rho_\infty$ is 
less than about 2. Their model consists of two layers of width $\delta$
with adjacent half spaces. The magnetic field is tangential to the boundaries
and asymmetric to the central boundary. The flow is parallel to the magnetic
field and symmetric with respect to the central boundary. Keeping the plasma
temperature constant throughout, there are besides the temperature six
parameters: magnetic field, flow speed and plasma density, each inside the
inner layers and in the half space, which can be varied. The unperturbed
pressure balance of the current sheet reduces the number of independent
parameters by one.

In view of this number of model parameters, we have reexamined
\citeauthor{Lee:etal:1988}'s dispersion relation (eqs.(34) and (35) of
\citeauthor{Lee:etal:1988}, \citeyear{Lee:etal:1988}) in the parameter range
we find most appropriate for our observations. We found above a half-width of
the streamer of about $\delta\simeq 0.2 \;R_\odot$ and a typical wave number
of $k=2\pi/2.5 \;R_\odot^{-1}$ for the wave structure which yields
$k\delta=0.5$. Assuming the magnetic field in the streamer
was half the outside value, the temperature was about 1 MK, and taking into 
account the pressure balance (equ.(10) in \citeauthor{Lee:etal:1988}, \citeyear{Lee:etal:1988})
\[
 \rho_p=\rho_{\infty}\left[\frac{T_\infty}{T_p}+\frac{V_{A\infty}^2}{2RT_p}\left(1-
 \frac{B_p^2}{B_\infty^2}\right)\right],
\]
the observed $\rho_s/\rho_\infty=1.5$ results in $V_{A\infty}\approx100\;\mathrm{km s^{-1}}$, 
and the plasma-$\beta$ then adjusts to values of 1.5 outside and 9 inside the
streamer.

In \citeauthor{Lee:etal:1988}'s \citeyearpar{Lee:etal:1988}
work, large constant growth rates are obtained for $k\delta>2$ for the both
kink and sausage mode. They are probably due to the sharp step-like gradients
between the piece-wise constant model profiles assumed. Numerical simulations
with more realistic profiles \cite[e.g.,][]{Zaliznyak:etal:2003} yield maximum
growth rates well below $k\delta\simeq 1$. From these dispersion relations it
is found that the phase speed does not vary drastically with wave number and
asymptotically in the limit $k\delta\rightarrow\infty$ becomes
\[
 V_\phi=V_\infty + \frac{\rho_s}{\rho_s+\rho_\infty}\Delta V_s
\]
where $\rho_\infty$ and $V_\infty$ are the density and flow speed outside the
streamer, $\rho_s$ and $V_\infty+\Delta V_s$ are the same parameters inside
the streamer. Note that different from \citet{Lee:etal:1988}, we here assume a wake model
for the streamer so that $\Delta V_s$ is negative. In order
to reach the observed phase velocities of $V_\phi\simeq400\dots350\;\mathrm{km s^{-1}}$, 
the choice of values for $V_\infty$ and $\Delta V_s$ is considerably constrained.
The set of values for which we
find a reasonable agreement with the observations is $V_\infty=500\;\mathrm{km
s^{-1}}$ and $\Delta V_s= -150\;\mathrm{km s^{-1}}=-1.5V_{A\infty}$.

In Fig.~\ref{fig:KinkDisp} for different values of the density ratio $\rho_s/\rho_\infty$, 
we show the threshold values for $\Delta V_s$, growth rate and phase speed for 
$V_\infty=500\;\mathrm{km s^{-1}}$ and $\Delta V_s=-150\;\mathrm{km s^{-1}}$.
From the threshold diagram we conclude that waves with $k\delta\simeq$ 0.3 to
0.5 can readily be excited by a velocity shear $\Delta V_s$ not much above
$V_{A\infty}$. We have to keep in mind that the Alfv\'en velocity is reduced
inside the streamer layer by a factor two. The adopted value 
of $|\Delta V_s|=1.5V_{A\infty}$ for the growth rate and phase speed below is just
above the threshold. If the threshold is exceeded, the
largest growth rates for the kink mode occur at $k\delta> 0.5$ and are
achieved for the density ratio $\rho_s/\rho_\infty$ between 1.5 and 3. For
larger density ratios, the sausage mode grows faster. Concerning
the observed density ratio about 1.5, the sausage mode is stable and has a zero
growth rate. Therefore, only the kink mode was observed. For $k\delta$ much
larger than 0.5, we believe that the analytical dispersion based on a model
with sharp, step-like gradients overestimates the growth rate. 

The phase speed
of the kink mode for the observed density ratio around 1.5 and the observed
wave number $k\delta\simeq 0.5$ is $V_\phi=3.8\;V_{A\infty} =380\;\mathrm{km
  s^{-1}}$. This comes again close to what we observed. Note that each
unstable kink and sausage mode is a coalescence of an up- and downshifted
propagating MHD boundary wave so that the phase speed splits for low wave
numbers where the mode becomes stable. When unstable, the two waves merge to a
pair with complex conjugate frequencies and a unique phase speed.

The growth rates rise quite steeply once the threshold is exceeded and a value
of $\gamma=0.2 kV_{A\inf}$ seems not unrealistic. However, we have to keep in
mind that the instability occurs in an inhomogeneous environment with fixed
boundary conditions on the sunward side of the streamer. At this side
the wave amplitude can only have noise level. Under these conditions, the
instability becomes convective instead of showing temporal growth.

The growth scale which we can expect is $V_\phi/\gamma=V_\phi/(0.2\,kV_{A\infty})$.
With the parameters used in our model, $V_\phi=380\;\mathrm{km s^{-1}}$ and
$V_{A\infty}=100\;\mathrm{km s^{-1}}$, we estimate the e-folding growth
scale to $3.8/0.2 \cdot \lambda/2\pi=3\lambda$. From the increase of the
observed wave amplitude apparent from the streamer traces
in Figure~\ref{fig:c2c3all}, this number does not seem unrealistic.

\section{Conclusions and discussion}

We present for the first time the observation of an oscillating streamer
probably driven by the streaming kink instability, a mode of the MHD
Kelvin-Helmholtz instability. The streamer presented 
a period of about 70 to 80 minutes, and its wavelength increased from
2 to $3\;R_\odot$ in about 1.5 hours. The phase speed of the oscillation
varied from $406\pm20$ to $356\pm31\;\mathrm{km s^{-1}}$ for different crests and trough.
These derived properties of the wave structure places constraints on
the plasma environment in which the instability took place.
We find a low Alfv\'en speed of about 100~$\mathrm{km s^{-1}}$
and a high flow speed of about 500~$\mathrm{km s^{-1}}$ outside the streamer.
It also turns out that the flow speed within the streamer was 
150~$\mathrm{km s^{-1}}$ lower.

In previous studies, oscillations of a streamer have 
been observed following interaction with a fast CME \citep{Chen:etal:2010,
Chen:etal:2011}. The phenomenon was termed as a ``streamer wave'' by the
authors. A distinct characteristic of their observation was the decrease of
the oscillation amplitude with time (see their Figure 4) due
to the convection of the energy with the outward propagation of the wave.
In the event reported here, no CME was detected in the LASCO field of view.
Also, the coronagraph observations by STEREO A and B did not show any CME
activity close to the streamer. Another distinction from the previously
reported oscillating streamer, within the same distance range, was the increasing 
amplitude of the wave with time. The streamer also
displayed a pronounced increase in amplitude when it convected to larger
heliocentric distances. This is evidence that the wave motion was excited by an instability.

Our assumptions are that the streamer is a 1D layer with plasma flow
and magnetic field mutually parallel and normal to the layer gradient
according to the model invoked by \citet{Lee:etal:1988}.
The plasma velocity is symmetrically distributed, the magnetic field varies
asymmetrical and the plasma temperature is constant throughout.
We do not distinguish between the widths of the velocity shear, the magnetic
field shear, and the scale of the density variation.
With these model assumptions we could reproduce many features of our
observation.

\begin{figure} 
\centering
\vbox{
\includegraphics[width=9.5cm, height=9.5cm]{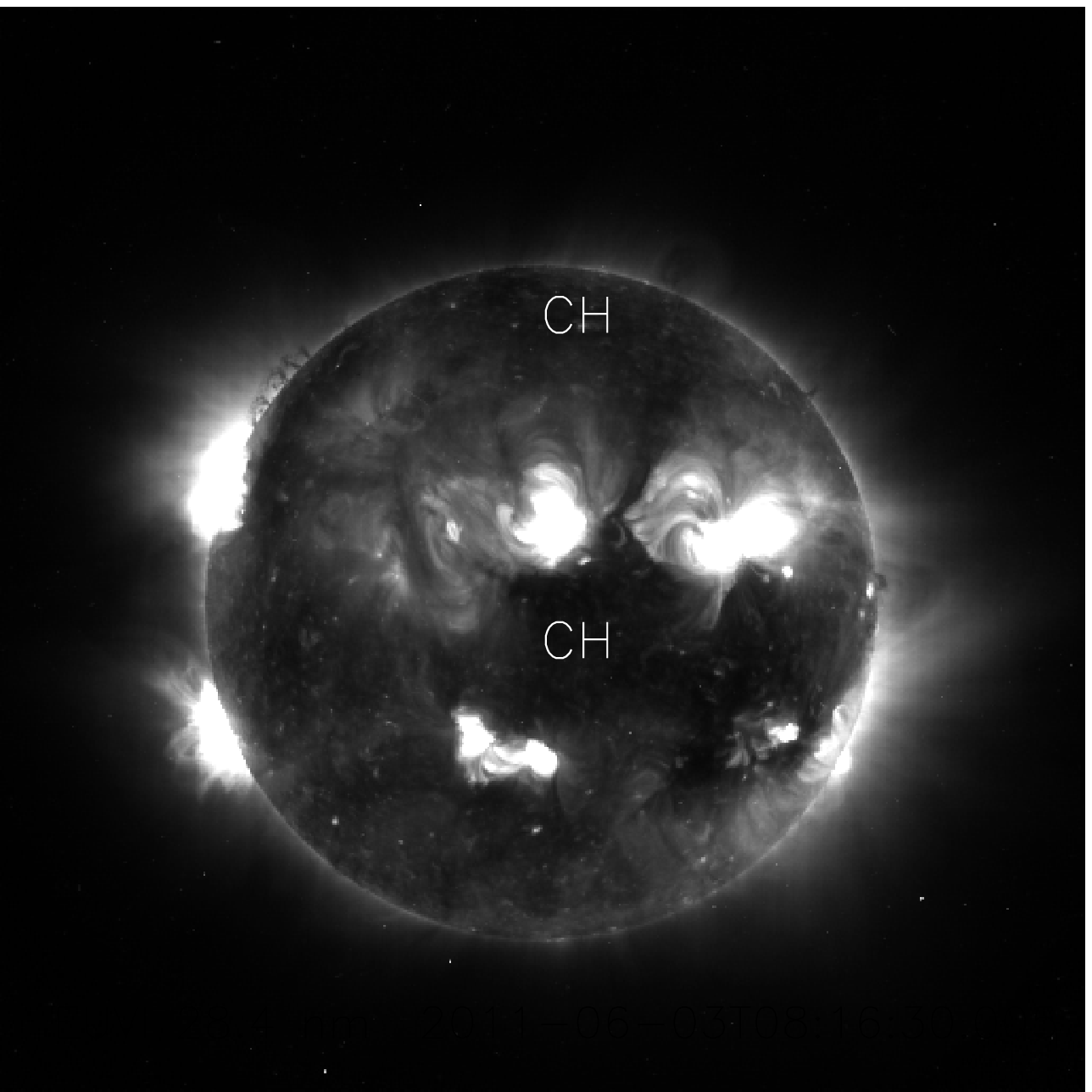}
\includegraphics[width=10.cm, height=10.cm]{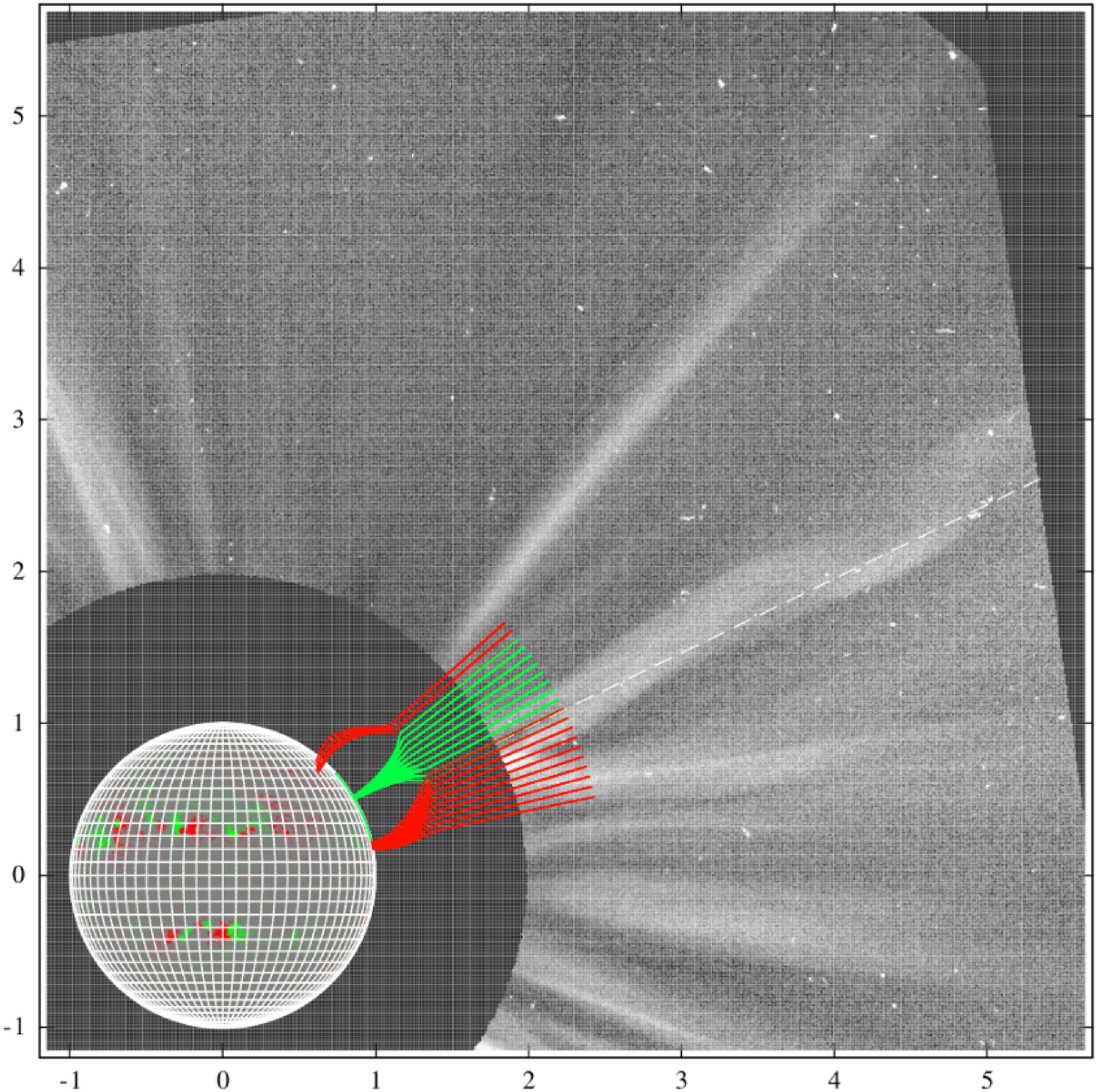}}
\caption{Upper panel: the EUV image observed at 28.4 nm by STEREO A/EUVI. 
On June 3 2011, STEREO A
was located about 90-degree west of the Earth. Two coronal holes above
and below the latitude of the wave-like feature are indicated.
Lower panel: the extrapolated coronal field lines from HMI synoptic map for
Carrington rotation 2110 using PFSS model. Red lines indicate inward open field lines,
and green lines represent outward open field lines. The spherical source surface
is set at $1.5\;R_\odot$.}
\label{fig:spatial_context}
\end{figure}

\begin{figure} 
\centering
\includegraphics[width=17.cm, height=20.cm]{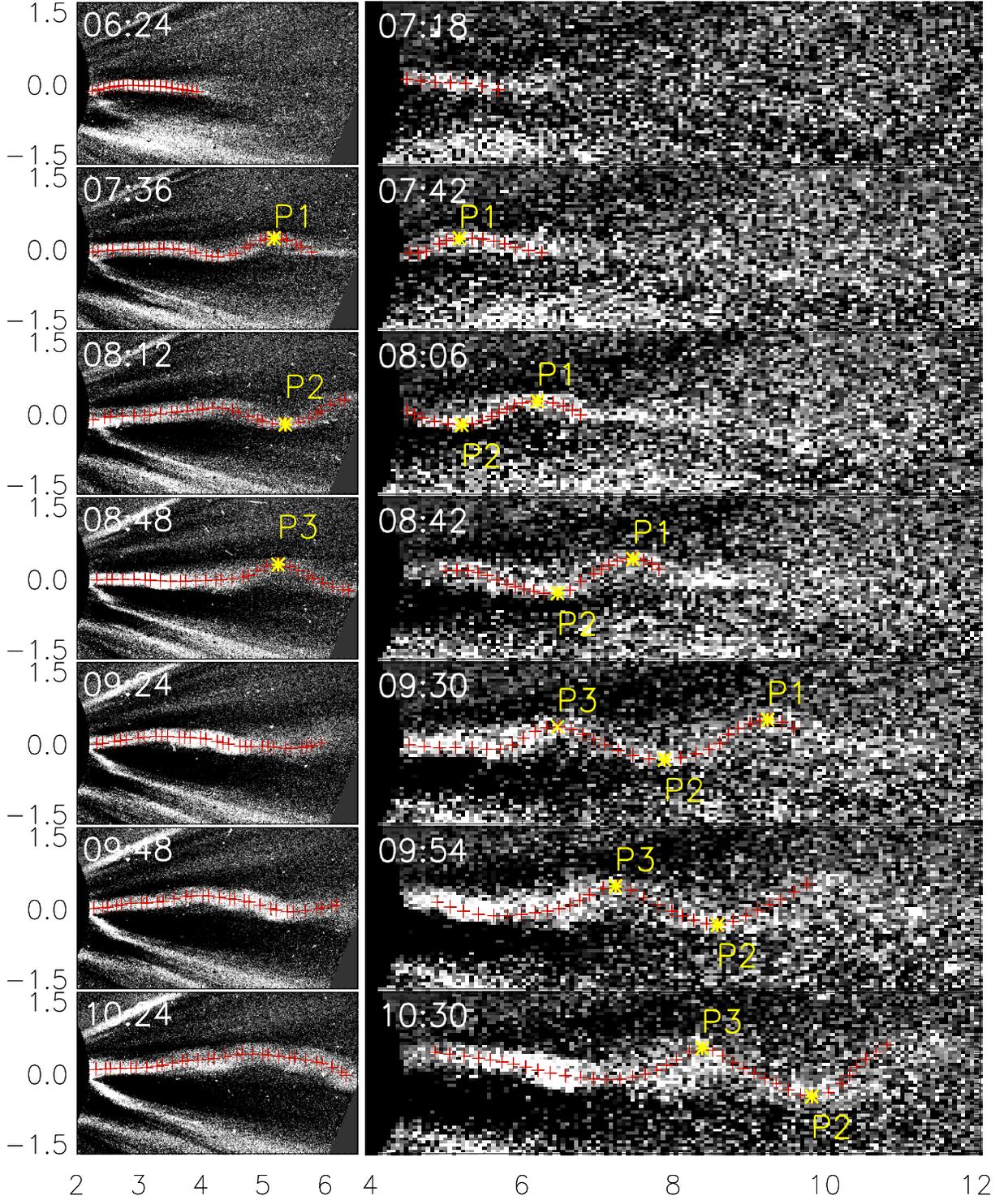}
\caption{Some selection of the traced streamer in C2 and C3 images. The streamer 
was rotated clockwise by $27.5^{\circ}$ to be horizontally located. The traced 
streamer at different instances is delineated by red plus signs. The first crest 
is indicated by P1, the first trough by P2, and the second crest by P3. For the 
context of the streamer, a strip from $-1.5\;R_\odot$ to $1.5\;R_\odot$ in 
north-south direction, from $2\;R_\odot$ to $6.5\;R_\odot$ in east-west direction 
was cropped in C2 images. In C3 images, the range in the east-west direction is 
from $4\;R_\odot$ to $12\;R_\odot$.}
\label{fig:c2c3trace}
\end{figure}

\begin{figure} 
\centering
\includegraphics[width=15.cm, height=20.cm]{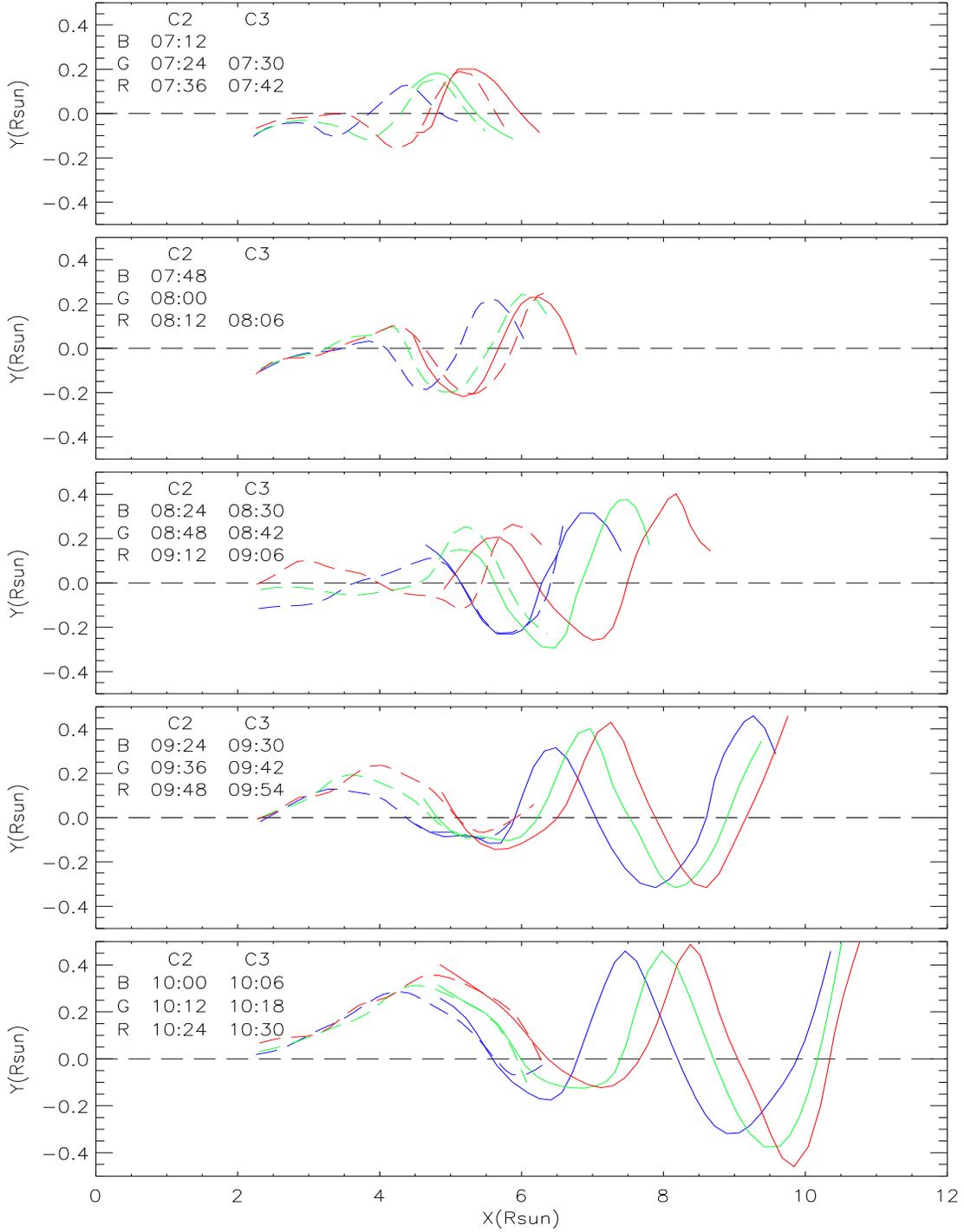}
\caption{The traced streamer in all the frames observed from 07:12~UT to 10:30~UT
by C2 and C3. The streamer within C2 FOV is indicated by long dashed lines, and 
within C3 FOV by solid lines. The streamer traced at different times are shown
in red (R), green (G), and blue (B) color. Those with observational time by C2 
and C3 within 6 minutes are displayed with the same color.}
\label{fig:c2c3all}
\end{figure}

\begin{figure} 
\centering
\includegraphics[width=17.cm, height=17.cm]{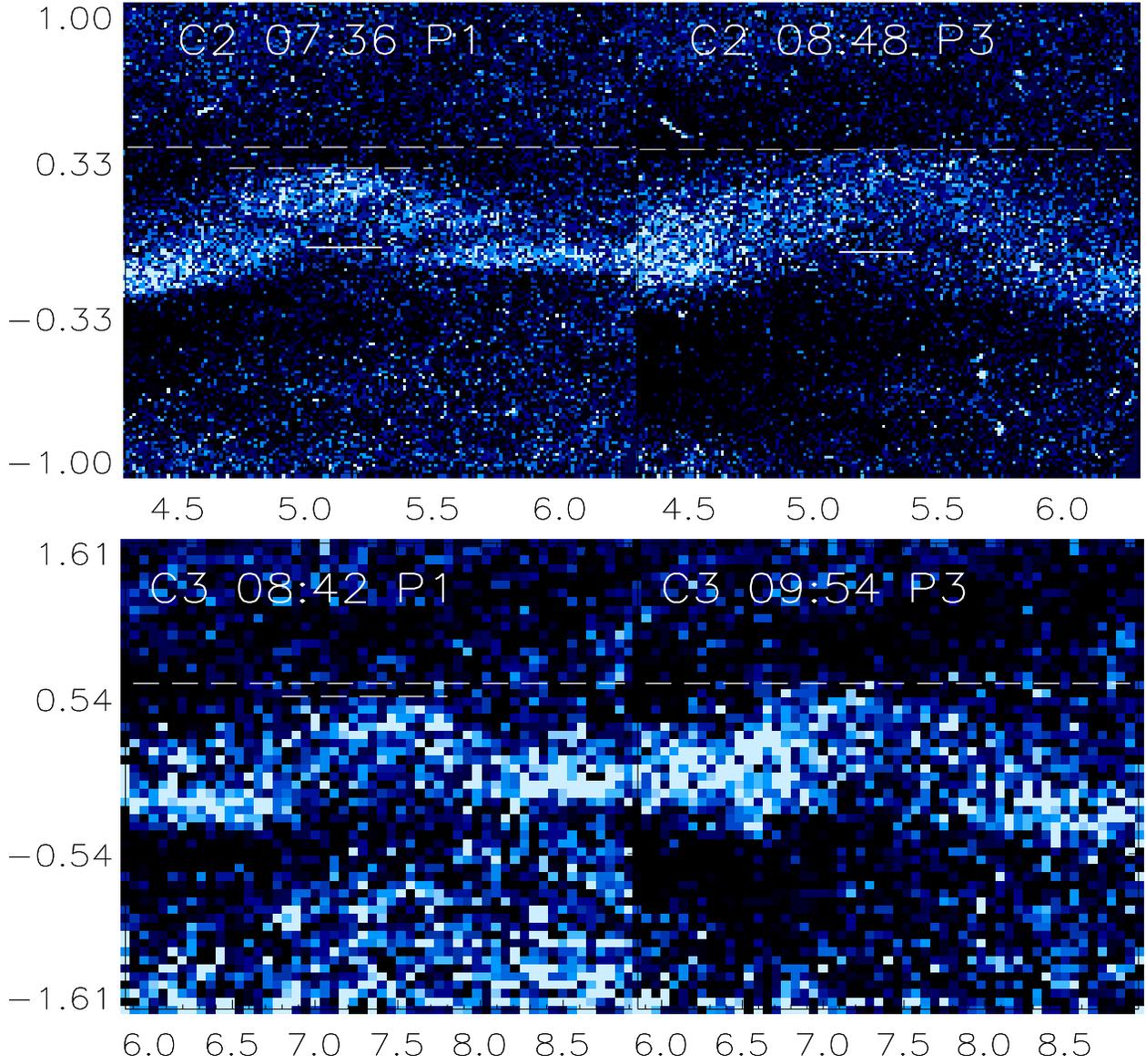}
\caption{Segments of the streamer around the crests P1 and P3 at 
different heliocentric distances.
Dashed lines mark the upper boundary of the crests P1 and P3, the
solid bars indicate their lower boundary. The $X$ and $Y$ axes are in units of 
$R_{\odot}$.}
\label{fig:amplitude}
\end{figure}

\begin{figure} 
\centering
\vbox{
\includegraphics[bb=100 70 480 300, clip=true, height=5.2cm]{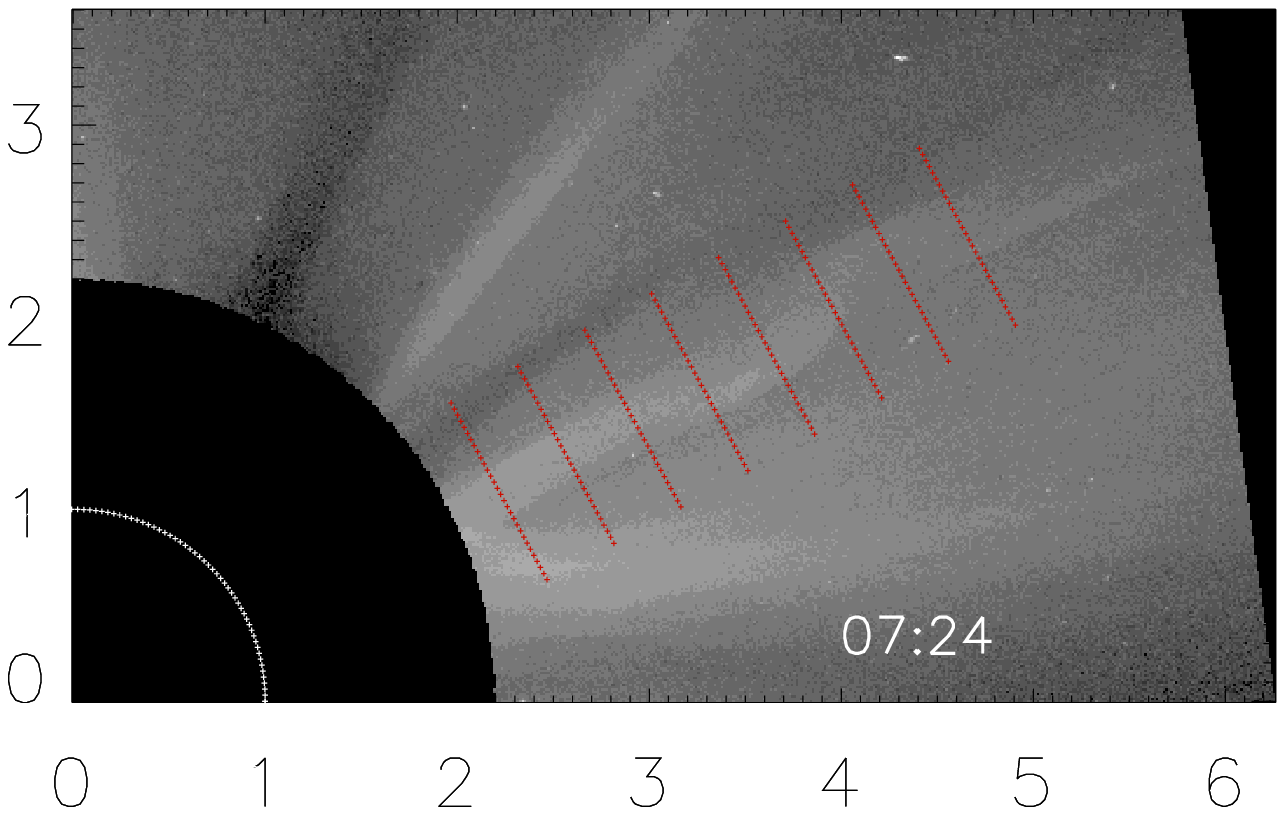}
\includegraphics[width=9.cm, height=5.3cm]{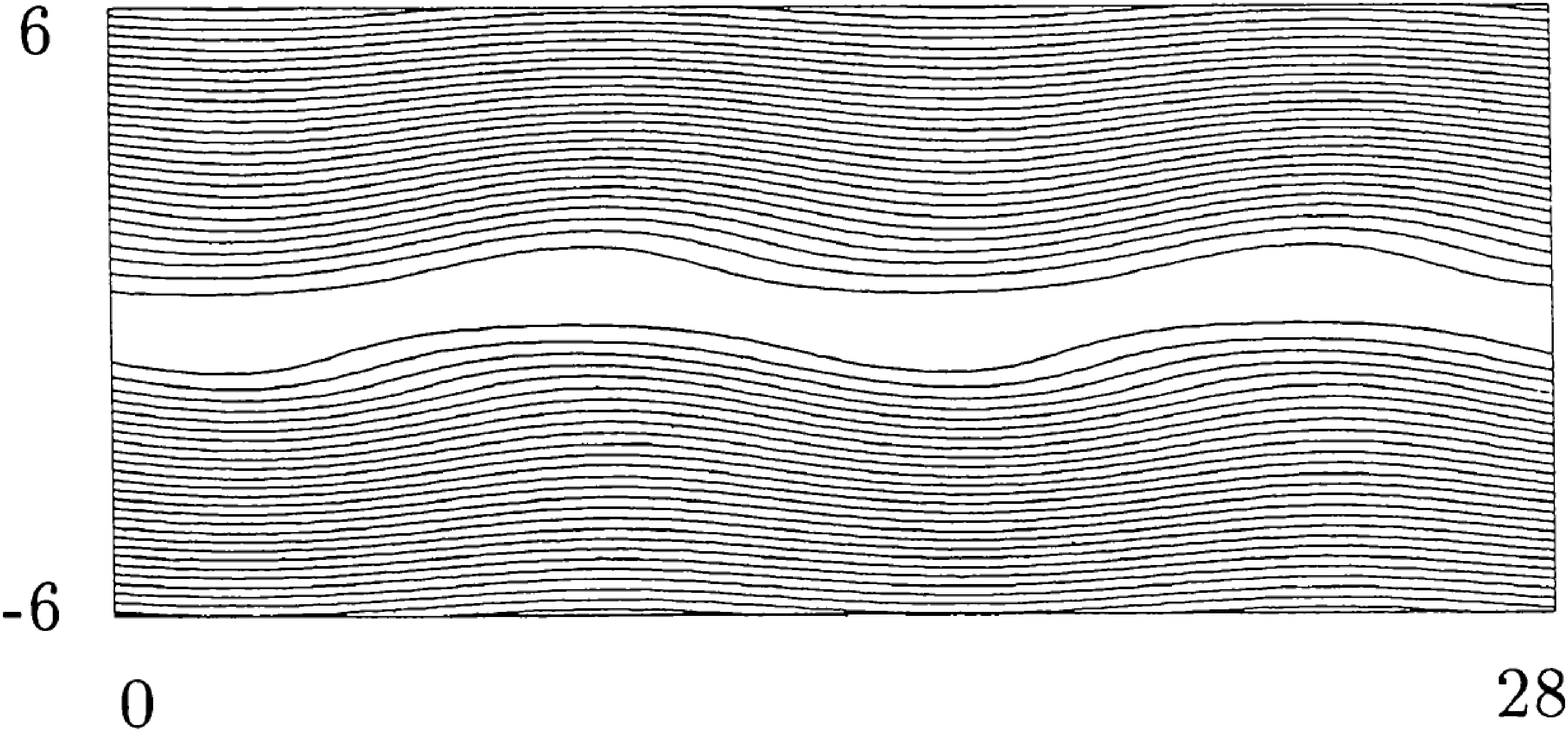}}
\caption{(Upper) A cut of the C2 image at 07:24~UT in which the streamer started to
show clear oscillations. Eight short red lines mark the positions where the brightness
distribution accross the streamer is measured. The $X$ and $Y$ axes are in units of 
$R_{\odot}$. (Lower) The magnetic field configuration of the streaming kink instability
adopted from Figure~1 in \citet{Wang:etal:1988}.  The $X$ and $Y$ axes are in units of 
the half thickness of the current layer.}
\label{fig:ROI_den}
\end{figure}

\begin{figure} 
\centering
\includegraphics[width=16.cm, height=10.cm]{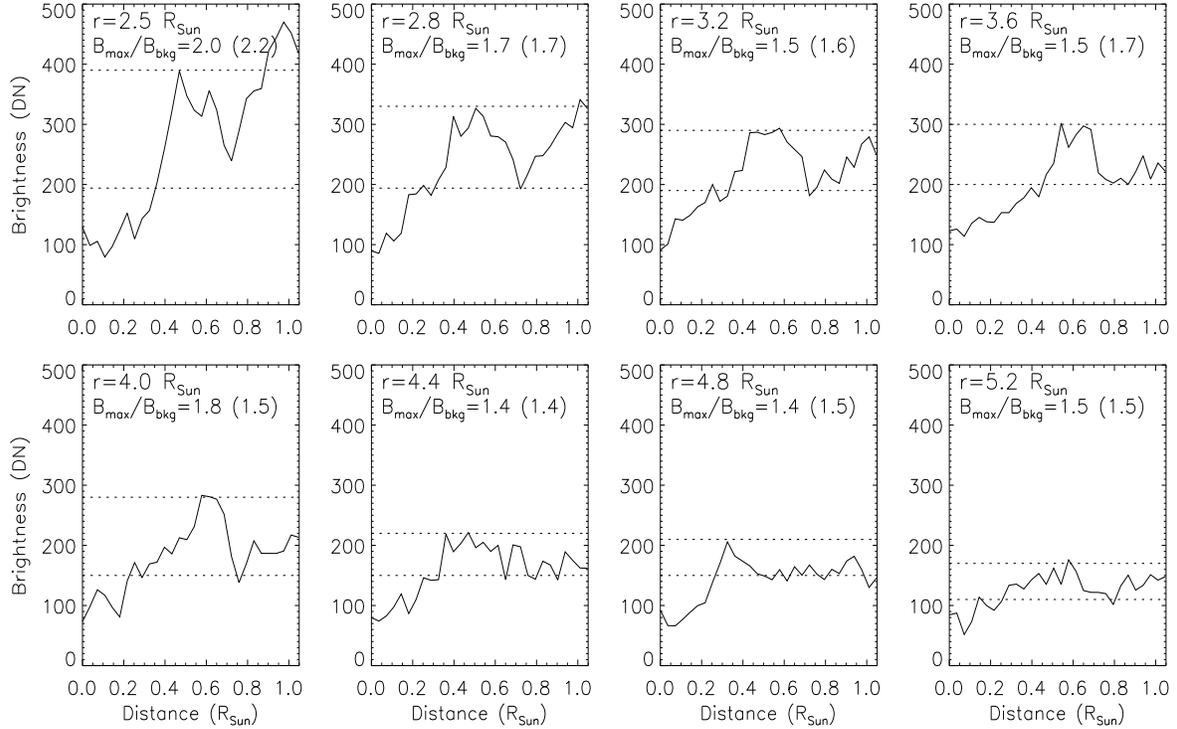}
\caption{The brightness distribution across the streamer at 07:24~UT along the
short red lines in Figure~\ref{fig:ROI_den} at different heliocentric distances. The distance
and the corresponding ratio of the maximal to the background brightness is
indicated in the upper position of each panel. Two different
background models for the stray-light and F corona subtraction, that is the monthly 
minimum and the yearly minimum, are applied. The ratio using the yearly minimum 
background is indicated in the parentheses in each panel. Two parallel dotted lines
represent the maximal and the background brightness. The distance along the $X$ axis 
is measured from the upper to lower positions on the short red line.}
\label{fig:den_ratio}
\end{figure}

\begin{figure} 
\centering
\includegraphics[width=15.cm, height=14.cm]{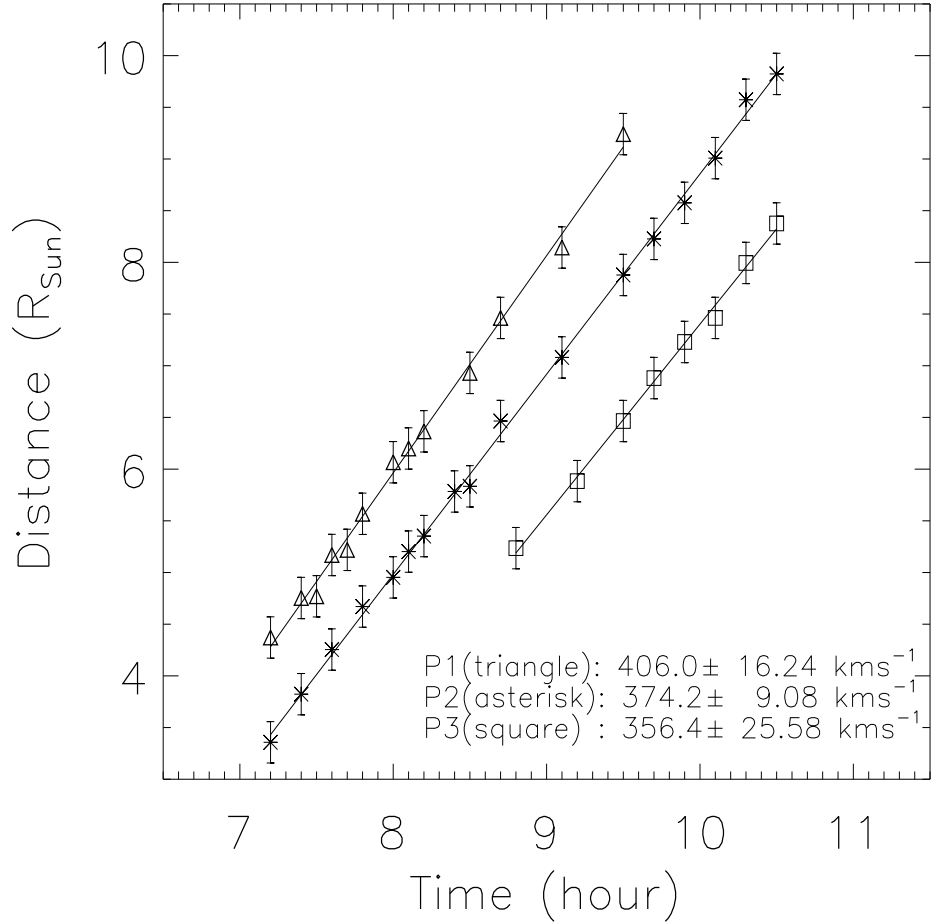}
\caption{The distance-time plot of P1, P2, and P3. Linear fits indicated by
solid lines are applied to derive their corresponding phase speeds, which are
406, 374 and $356\;\mathrm{km s^{-1}}$. The errorbars are the assumed 
$\delta=\pm0.2\;R_\odot$ uncertainties. The propagation of this error to the phase speeds
are indicated in the right corner as the 1-sigma uncertainty of the fitted slope.}
\label{fig:speed}
\end{figure}

\begin{figure} 
\centering
\includegraphics[bb=2 2 440 420, clip, width=7cm]{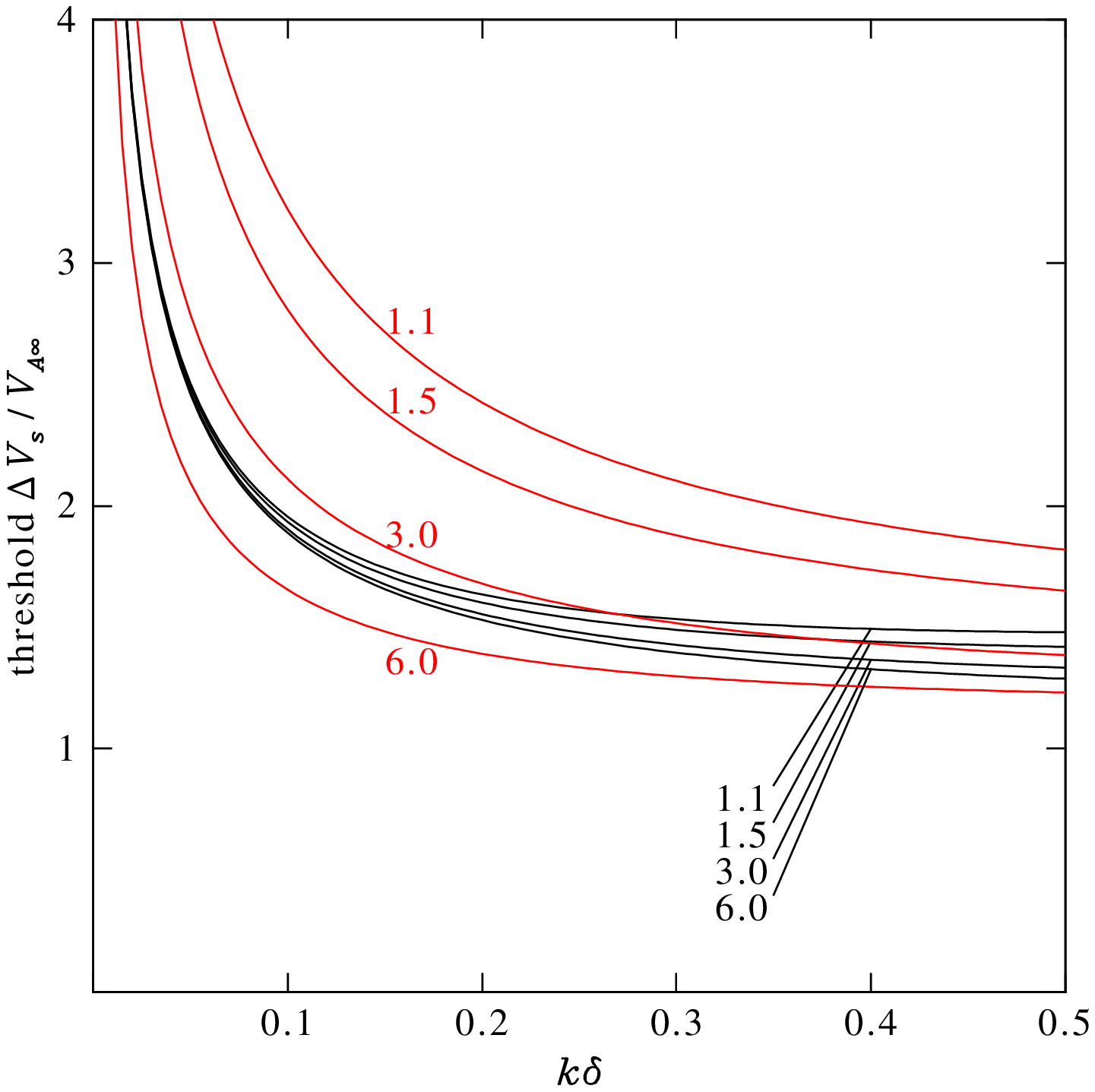}\\
\includegraphics[bb=2 2 440 420, clip, width=7cm]{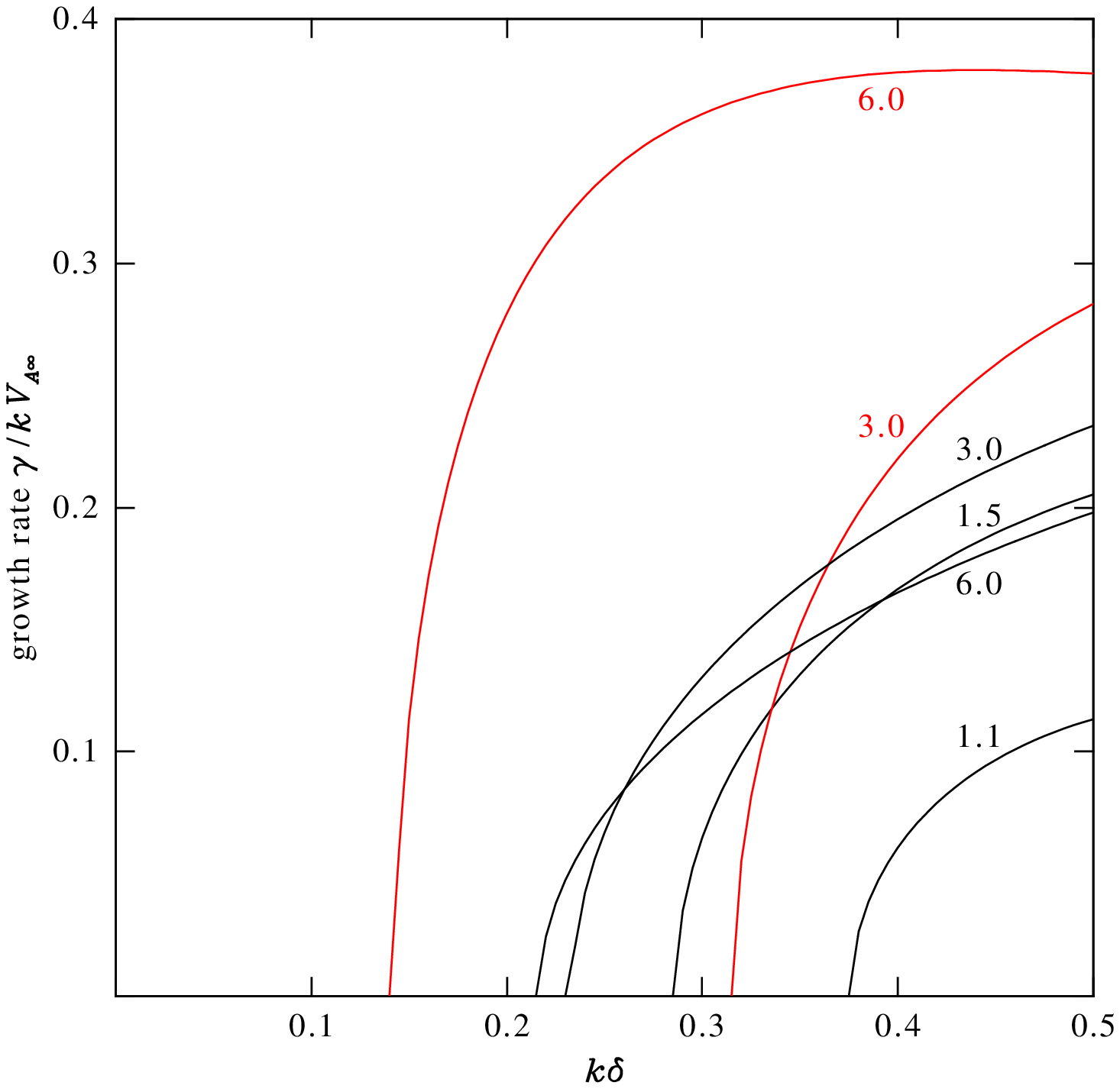}\\
\includegraphics[bb=2 2 440 420, clip, width=7cm]{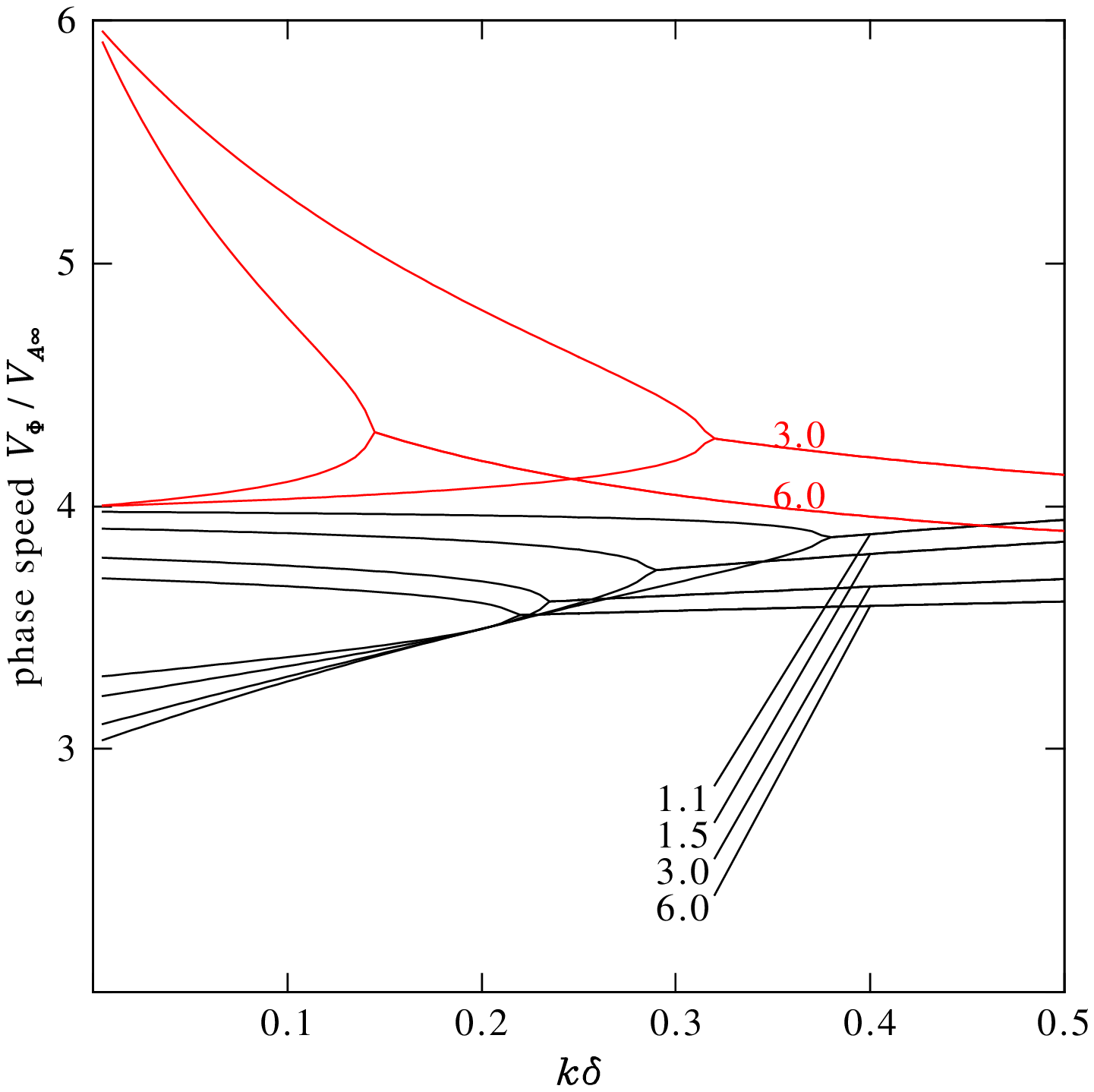}
\caption{Normalized velocity shear threshold, growth rate and phase speed
versus wave number for different density ratios $\rho_s/\rho_\infty$ from
the dispersion equation for the kink(black) and sausage(red) mode of
\citep{Lee:etal:1988}.
The curves are labelled according to the density ratio $\rho_s/\rho_\infty$.
For the ratios 1.5 and 1.1, the sausage mode is stable in the wavelength
range shown.
Growth rate and phase speed were calculated for $\Delta V_s=-1.5\;V_{A\infty}$.
  Other parameters are given in the text.}
\label{fig:KinkDisp}
\end{figure}

In order to reproduce the observed phase speed we had to invoke a high flow
speed and a low Alfv\'en speed outside of the streamer.
Typically assumed values for slow solar wind in the distance
range from 8 to $10\;R_\odot$ are $V_{A\infty}\simeq 200\;\mathrm{km s^{-1}}$
\citep{Warmuth:Mann:2005} and $V_\infty\simeq 200\;\mathrm{km s^{-1}}$
\citep[e.g.,][]{Sheeley:etal:1997}.
The latter authors determined $V_\infty$ from the velocity of density
fluctuations drifting off streamer tips. They also show that the speed of
individual density blobs scatters enormously around the above quoted average
and may reach more than $600\;\mathrm{km s^{-1}}$ in some cases.
The value of $V_\infty= 500\;\mathrm{km s^{-1}}$ which we
conclude from our observations is therefore not unrealistic. Moreover, at the
time of our observations, the streamer was sandwiched between two nearby
coronal holes to the north and south of the streamer (see Figure~\ref{fig:spatial_context}). 
It does not seem
impossible, that the fast solar wind from these coronal holes had some impact
on the flow speed $V_\infty$ in the vicinity of the streamer.

In our analysis, we also had to assume a low Alfv\'en speed
$V_{A\infty}$ to reproduce the observed phase speed of 350 to $400\;\mathrm{km
s^{-1}}$. Our value $V_{A\infty}=100\;\mathrm{km s^{-1}}$ is low but still
slightly higher than the value taken by \citet{Chen:etal:2009} in their
simulations of a coronal streamer.
Higher values for $V_{A\infty}$ require even larger flow speeds $V_\infty$ and also a
larger velocity shear magnitude $|\Delta V_s|$ inside the streamer. From our estimates,
$\Delta V_s=-1.5 V_{A\infty}$ just exceeds the threshold condition. For a
larger $\Delta V_s$ and also for density ratios $\rho_s/\rho_\infty>3$,
the kink instability would be superseded by the sausage mode. We believe that
these special conditions are the reason why the kink mode is only seldom
observed with coronal streamers compared to the sausage mode. This alternative
mode manifests itself in the formation of density fluctuations and plasmoids
which are more commonly observed in coronal streamers.

The streamer is also special in that its cusp is probably lower than
for conventional streamers. We have tried to construct the magnetic field
near the streamer by a PFSS model. For mathematical reasons, these models
envoke a spherical outer boundary at some distance $r_{SS}$ where the field
becomes entirely radial. Typically, $r_{SS}$ is chosen between 2 and 2.5
$R_\odot$, but for the real coronal magnetic field, the streamer cusps
do not need to have $r_{SS}$ at the same height. Modelling the field in the
vicinity of the streamer we found that its current sheet is mapped into
heliosphere at exactly the observed latitude if $r_{SS}$ is given a
value of less than 1.6 $R_\odot$.

The alternative possiblity is a pseudostreamer configuration
\citep{Wang:etal:2007} for the observed oscillating streamer. In the case
of the pseudostreamer, the initial magnetic field has the same direction
in different layers. We have gone through the formula in \citet{Lee:etal:1988}.
to check how the dispersion equation and the shape of a streamer would
change when the configuration of magnetic field varies from the case of 
helmet streamer to the case of pseudostreamer. Although the details of 
the derivation is beyond the scope of this paper and not presented here, 
surprisingly, we find that the dispersion
equation and the related instability threshold, growth rate, phase speed, and
the shape of the streamer do not change. The only difference between the 
resulting magnetic field is its direction. For helmet streamers, its direction
is opposite on two sides of the streamer axis, whilst for pseudostreamers, the
direction of magnetic field is the same.

\acknowledgements

We thank the anonymous referee for his insightful comments and careful reading of  
our paper.
We also benefited from the valuable discussions with Yao Chen, Jiansen He and Hui Tian.
SoHO and STEREO are projects of international 
cooperation between ESA and NASA. The SECCHI data are produced by an international 
consortium of NRL, LMSAL, and NASA GSFC
(USA), RAL, and U. Birmingham (UK), MPS (Germany), CSL (Belgium), IOTA, and IAS
(France). SDO is a mission of NASAs Living With a Star Program.
L.F. and W.Q.G. are supported by by MSTC Program 2011CB811402, NSFC under grant 
11003047, 11233008, 11273065 and by Grand 
BK2012889. L.F. also acknowledges the Key Laboratory of Dark Matter and Space 
Astronomy, CAS, for financial support.
The work of B.I. is supported by DLR contract 50 OC 1301.

\end{document}